\documentclass[aps,prl,twocolumn,superscriptaddress,noshowpacs]{revtex4-2}

\usepackage{eqnarray}
\usepackage{amsmath,amsfonts,amssymb}

\DeclareMathOperator{\cn}{cn}
\DeclareMathOperator{\dn}{dn}
\DeclareMathOperator{\sn}{sn}
\usepackage{graphicx}
\usepackage{bm}
\usepackage{color}
\usepackage{physics}

\begin{document}

\title{Beam canalization by a non-Abelian gauge field}

\author{Olha Bahrova}
\email{olha.bahrova@uca.fr}
\affiliation{Institut Pascal, PHOTON-N2, Universit\'e Clermont Auvergne, CNRS, Clermont INP,  F-63000 Clermont-Ferrand, France}
\affiliation{B. Verkin Institute for Low Temperature Physics and
Engineering of the National Academy of Sciences of Ukraine, 47
Nauky Ave., Kharkiv 61103, Ukraine}

\author{Jiahao Ren}
\affiliation{Division of Physics and Applied Physics, School of Physical and Mathematical Sciences, Nanyang Technological University, Singapore, Singapore}

\author{Feng Jin}
\affiliation{Division of Physics and Applied Physics, School of Physical and Mathematical Sciences, Nanyang Technological University, Singapore, Singapore}

\author{Rui Su}
\affiliation{Division of Physics and Applied Physics, School of Physical and Mathematical Sciences, Nanyang Technological University, Singapore, Singapore}
\affiliation{School of Electrical and Electronic Engineering, Nanyang Technological University, Singapore, Singapore}

\author{Guillaume Malpuech}
\email{guillaume.malpuech@uca.fr}
\affiliation{Institut Pascal, PHOTON-N2, Universit\'e Clermont Auvergne, CNRS, Clermont INP,  F-63000 Clermont-Ferrand, France}

\author{Dmitry Solnyshkov}
\affiliation{Institut Pascal, PHOTON-N2, Universit\'e Clermont Auvergne, CNRS, Clermont INP,  F-63000 Clermont-Ferrand, France}
\affiliation{Institut Universitaire de France (IUF), 75231 Paris, France}

\begin{abstract}
Hyperbolic and quasi-flat isofrequency contours (IFCs) are used for beam canalization and can be created by tilted Dirac points in photonic systems. Dirac points in microcavities are generated by the combination of transverse-electric/transverse-magnetic splitting and linear birefringence. We show that the canalization is here strongly assisted by the coupling between the spatial dynamics and polarization pseudospin precession. This dynamics 
is well described analytically and numerically as the action of a non-Abelian gauge field acting on emergent charges (spin current).
 We demonstrate a ten-fold enhancement of the canalization for a Gaussian beam by the gauge field, as compared to a description based solely on the group velocity associated with the IFCs.
\end{abstract}

\maketitle

The implementation of photonic systems showing hyperbolic isofrequency contours (IFC) has been a very active topic of research in the last decade~\cite{ferrari2015hyperbolic,guo2020hyperbolic,lee2022hyperbolic,wang2024planar,Ermolaev2026}. This peculiar situation takes place for surface modes in bi-anisotropic media, when the permittivity along the optical axis and perpendicular to it are of opposite sign. It can be realized using so-called hyperbolic metamaterials \cite{poddubny2013hyperbolic,high2015visible} or by using guided polariton modes near the plasmonic, phononic or excitonic \cite{ruta2023hyperbolic} resonances of 2D materials yielding the concept of hyperbolic polaritons \cite{hu2020topological}. Hyperbolic photonic modes can also be present and used in pure dielectric photonic crystals \cite{li2003evaluation,kavokin2005negative},
for instance near the M-point of a square lattice \cite{arlandis2012mesoscopic}. A very intriguing case occurs exactly at the transition between hyperbolic to parabolic IFCs when the iso-frequency lines become straight, giving rise to a quasi-perfect beam collimation called "canalization" \cite{belov2005canalization,hu2020topological,duan2025canalization}, which has been reported for phonon-polariton modes, in the deep infrared region of the spectrum.

On the other hand, despite of their apparent simplicity, planar microcavities~\cite{kavokin2017microcavities} imposed themselves as paradigmatic systems to engineer 2D photonic modes demonstrating topological singularities such as Dirac points, and topological transitions~\cite{kavokin2005,leyder2007,terccas2014non,nalitov2015polariton,klembt2018exciton,richter2019voigt,Rechcinska_Science2019,gianfrate2020measurement,ren2021nontrivial,krol2021observation,polimeno2021tuning,spencer2021spin,long2022helical,krol2022annihilation,liang2024polariton}.
One key ingredient behind this richness is the vicinity between modes of different polarizations and their coupling by various types of effective spin-orbit coupling (SOC). Another important ingredient is related to the possibility of filling cavities with semiconductors showing excitonic resonances which mix with photonic modes realizing cavity exciton-polaritons. This brings new properties such as strong spin-dependent non-linearities (spinor quantum fluid physics) \cite{carusotto2013quantum}, sensitivity to external applied fields~\cite{polimeno2021tuning,Rechcinska_Science2019}, and the possibilities to undergo bosonic phase transitions such as Bose-Einstein condensation (BEC).

\begin{figure}
    \centering
    \includegraphics[width=\linewidth]{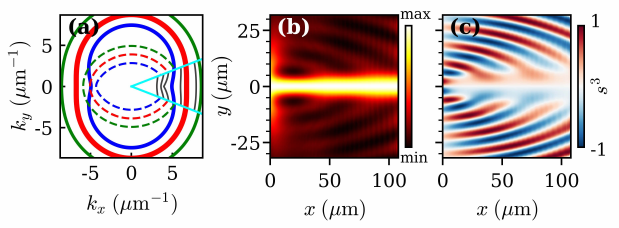}
    \caption{(a) Schematic representation of main types IFCs marked by thick curves: hyperbolic (blue), flat (red) and parabolic (green). The dashed curves shows the corresponding inner band IFCs. The gray curves plotted within the highlighted sector indicate additional IFCs.   (b) Total polariton intensity, obtained from the SE simulation, and normalized along the x axis manifesting pronounced canalization effect strongly enhanced by the present effective non-Abelian gauge field. (c) $s^3$ Stokes vector component.}
    \label{fig1}
\end{figure}

The most prominent SOC, which was described quasi-simultaneously in microcavities~\cite{kavokin2005} and in other systems~\cite{bliokh2008geometrodynamics}, is related to the transverse (transverse-electric/transverse-magnetic or TE-TM) nature of light modes. In microcavities, the light wavevector perpendicular to the mirrors is quantized whereas the in-plane motion is parametrized by a 2D wave vector $k$. The resulting quantized modes show a 2D parabolic dispersion, with a degeneracy at $k=0$ between the TE and TM modes which typically demonstrate different effective masses. The "double-headed" vector of polarization of these modes is characterized by a winding number 2 with respect to the polar angle of the wave vector and  a winding number 2 for the associated polarization pseudospin (which is an "ordinary" vector). This intrinsic winding is a key property to implement topologically non-trivial photonic modes \cite{nalitov2015polariton}. It was initially revealed through the observation of the Optical Spin Hall effect \cite{kavokin2005,leyder2007} (OSHE), where a focused linearly-polarized normally incident pump excites both TE and TM circular IFCs. The resulting ballistic expansion is combined with a k-dependent pseudospin precession which leads to a spatial separation of circular polarization components. Linear birefringence modifies the band geometry, transforming the parabolic band touching into a pair of tilted Dirac points, each characterized by a pseudospin winding number 1 \cite{terccas2014non,gianfrate2020measurement}. The IFCs close to the Dirac point energy are highly anisotropic, showing a "camelback" 2-humped shape, passing from strongly hyperbolic to parabolic as shown in Fig.~\ref{fig1}(a). As in the other cases, this transition occurs via a quasi-flat IFC with a cancellation of the quadratic wavevector term, allowing one to observe a strong canalization effect~\cite{ren2026optically}. 
However, with respect to other implementations of hyperbolic IFCs, the associated mode polarization texture leads to a peculiar coupling between spin and real space dynamics.
 
The effective Hamiltonian describing the system can be mapped to the one of vectorial charges (spin-current) evolving in a non-Abelian gauge potential \cite{terccas2014non,yang2024non, song2025} whose dynamical action can be described by using a Yang-Mills theory for SU(2). When the exciting wavepacket is circularly polarized, the resulting dynamics is similar to Zitterbewegung, as experimentally observed in different platforms \cite{polimeno2021experimental,hasan2022wave,lovett2023observation,wen2024trembling}. The resulting non-Abelian analogue of the Lorenz force~\cite{jin2006su2} was recently shown to lead to topological transitions in localized photonic structures~\cite{torons,widmann2026}.
However, the effect of the polarization and of this Lorenz force on the canalization has not been studied so far.

In this work we consider a microcavity that could contain a semiconductor with strong excitonic resonances leading to the formation of strongly coupled cavity exciton-polaritons \cite{weisbuch1992observation,kavokin2017microcavities}. The presence of both TE-TM SOC and linear birefringence creates a pair of tilted Dirac cones in the mode dispersion. 
We show that when the quasi-flat IFCs occur close to the Dirac points, the canalization, which is not ideal due the $k^4$ contribution to the IFC, is combined with the OSHE, and the non-Abelian gauge field plays an important role, compressing the injected beam. The resulting canalization enhancement depends on the initial beam width and can reach a factor 10 for a narrow Gaussian beam.

\textit{The model.} We consider the effective Hamiltonian of photonic or polaritonic modes in a planar microcavity or a similar structure combining the TE-TM spin-orbit coupling with a constant birefringence~\cite{terccas2014non}:
\begin{eqnarray}
H_{\mathbf{k}}=\begin{pmatrix}
E_0+\frac{\hbar^2k^2}{2m}&\beta_0-\beta k^2 e^{-2i\varphi} \\
\beta_0-\beta k^2 e^{2i\varphi}& E_0+ \frac{\hbar^2k^2}{2m}
\label{Hameff}
\end{pmatrix} 
\end{eqnarray}
where $m=m_{TM}m_{TE}/(m_{TM}+m_{TE})$, $k=|\mathbf{k}|=\sqrt{k_x^2+k_y^2}$ is the in-plane wavevector ($k_x=k\cos{\varphi}$, $k_y=k\sin{\varphi}$.), $\beta$ is the TE-TM splitting strength and $\beta_0$ is the optical birefringence.
The two parabola cross at $k^0_x=\pm(\beta_0/\beta)^{1/2}$, giving rise to two tilted Dirac points (also called diabolical points in optics). In the vicinity of these points, the IFCs (Fig.~\ref{fig1}(a)) pass from hyperbolic to parabolic via the flat regime, where the canalization is observed~\cite{ren2026optically}. This is illustrated by a simulation based on the Hamiltonian~\eqref{Hameff} in real space representation, showing the intensity of a propagating beam as a function of coordinates (Fig.~\ref{fig1}(b)). The beam indeed remains quite tight. However, the canalization of the total intensity is accompanied by an alternating and anti-symmetric pattern of the circular polarization degree represented by the Stokes vector component $s^3$ (Fig.~\ref{fig1}(c)). It appears because the polarization state of the beam at the injection point does not match the eigenstates for all wave vectors contained in the beam.  A polarization precession is observed following the OSHE mechanism. The wave packet components with positive $k_y$
show a pseudo-spin precession in anti-phase with respect to the ones with negative $k_y$ because of the opposite sign of the polarisation splitting in k-space on opposite sides of the Dirac point. This pseudospin precession is accompanied by a Zitterbewegung-like oscillatory spatial motion being in anti-phase for the positive and negative $k_y$ components as well. Altogether, the resulting full wave packet behaves as a canalized beam, as we explain in details below.

Around each of these Dirac points, the Hamiltonian~\eqref{Hameff} can be rewritten \cite{terccas2014non,gianfrate2020measurement,polimeno2021experimental} as a Rashba-like Hamiltonian \cite{Rashba1984}
\begin{equation}
    \hat{H}_R=\frac{1}{2m}\hat{\bm{p}}^2+\alpha \bm{\sigma}\cdot\hat{\bm{p}}=\frac{1}{2m}\left(\hat{\bm{p}}+m\alpha \bm{\sigma}\right)^2-m\alpha^2\sigma^0
    \label{hamR}
\end{equation}
where $\mathbf{\sigma}$ is a vector of Pauli matrices, $\bm{p}=\hbar \bm{q}$ is the momentum, $\bm{q}=\bm{k}-\bm{k}^0$, $\alpha=\sqrt{\beta_0\beta/2}$.
This Hamiltonian can be interpreted as a general Hamiltonian of a matter field minimally-coupled with a Yang-Mills field~\cite{jin2006su2,Tokatly2008}:
\begin{equation}
    H_{YM}=\frac{1}{2m}\left(\hat{\bm{p}}-\eta\bm{A}^a\sigma^a\right)^2+\eta A_t^a\sigma^a
    \label{HYM}
\end{equation}
with two non-zero components of the emergent vector potential: $A_x^1=-m\alpha/\eta$, $A_y^2=-m\alpha/\eta$. Thanks to the non-Abelian nature of the gauge field, a constant vector potential results in a non-zero field strength tensor with components given by $F_{\mu\nu}^a=\partial_\mu A_\nu^a-\partial_\nu A_\mu^a-\eta\varepsilon^{abc}A_\nu^b A_\mu^c$, where $\mu,\nu$ span $(t,x,y,z)$. The nonzero components read: $F_{yx}^3=-F_{xy}^3=-m^2\alpha^2/\eta$.

\begin{figure}
    \centering
    \includegraphics[width=1.0\linewidth]{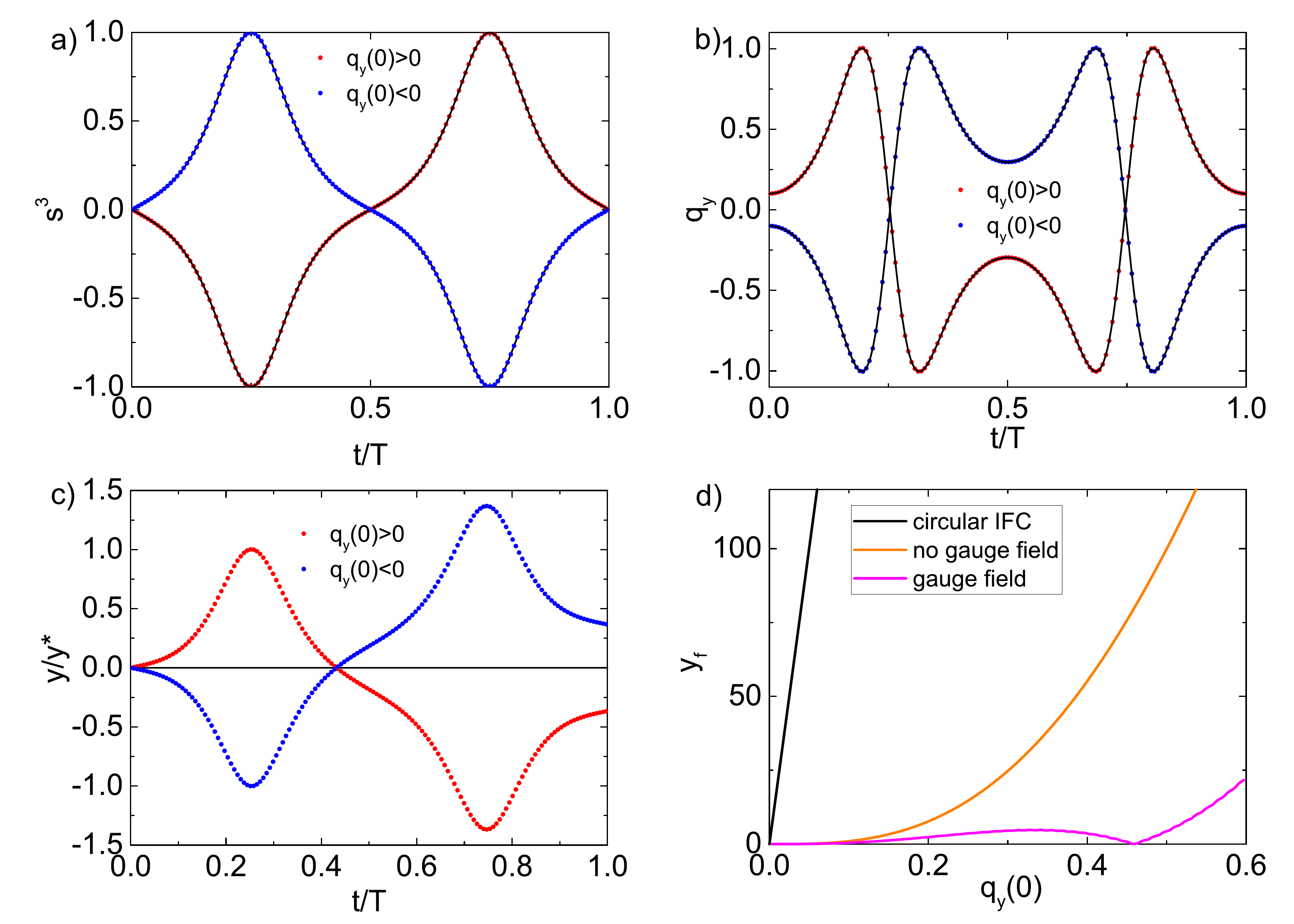}
    \caption{a) The spin projection $s^3(t)$ for a single period: numerical and analytical solutions. b) The $q_y(t)$ projection: numerical and analytical solutions. c) The transverse coordinate as a function of time $y(t)$ (numerics, $y^*$ is the oscillation amplitude). d) Final shift with and without the gauge field as a function of initial wave vector $q_y(0)$ (numerics). }
    \label{fig2}
\end{figure}

The force provided by the Yang-Mills field links a unified spin-current vector $\bm{J}$ and the field strength tensor $\bm{F}$. The equations of motion for the velocity $\bm{v}$ and spin $\bm{s}$ of a classical relativistic particle coupled to the Yang-Mills field read:
\begin{equation}
    m\,dv^\mu/d\tau=\bm{J}_\nu\cdot\bm{F}^{\mu\nu},\quad d\bm{s}/d\tau=-\eta\bm{A}_\mu\times\bm{J}^\mu
    \label{chrome}
\end{equation}
where $\bm{J}_\nu=\bm{s}v_\nu$ is the spin current. Since only the $F^3$ components are non-zero, an analogue of the Lorenz force can appear only for the $s^3$ pseudospin component (circular polarization current). This is why the non-zero projection of the $s^3$ Stokes vector component observed in Fig.~\ref{fig1}(c) is important. In what follows, we compute semiclassical trajectories of particles, which are supposed to be the components of a wavepacket (propagating beam), depending on their initial wavevector. The full beam is then described as a superposition of these trajectories. This approach gives an appropriate description when the interference effects do not play a strong role on the wave front profile, as confirmed by the direct solution of the spinor Schrödinger equation (Fig.~\ref{fig1}(b,c)).

The varying profiles of the IFCs require accounting for second-order terms in the kinetic energy. We use the Hamiltonian formalism to rewrite the equations of motion:
\begin{eqnarray}
    d\bm{s}/dt&=&-\eta\bm{A}_\mu\times\bm{J}^\mu\label{spidyn}\\
    dp^\mu/dt&=&\bm{J}_\nu\cdot\bm{F}^{\mu\nu}\label{Lorefor}\\
    dx^\mu/dt&=&\partial H/\partial p^\mu\label{spadyn}
\end{eqnarray}
The important difference with respect to purely linear Dirac Hamiltonian is that now the group velocity $dx^\mu/d\tau=\partial H/\partial p^\mu$ is not necessarily aligned with the momentum $\bm{p}$. The two components of this velocity read explicitly
\begin{eqnarray}
    \frac{dx}{dt}&=&\frac{\hbar(k_0+q_x)}{m}-\frac{\alpha q_x}{\hbar q}\\
    \frac{dy}{dt}&=&\frac{\hbar q_y}{m}-\frac{\alpha q_y}{\hbar q} \label{vgy}
\end{eqnarray}
where $q=\sqrt{q_x^2+q_y^2}$. The canalization by the group velocity profile, arising from the quadratic corrections to the energy, is ensured by the opposite signs of the two contributions to $dy/dt$, which cancel each other for $q=\alpha m/\hbar^2$. It means that the quasi-flat IFC passes through the point $q_x=\alpha m/\hbar^2$, $q_y=0$.

The key result of this work is that the gauge field provides an extra canalization. The pseudospin precession (Eq.~\eqref{spidyn}), arising via the OSHE mechanism,  creates a non-zero $s^3$ spin component with a sign depending on the sign of $q_y$ (as shown in Fig.~\ref{fig1}(c)). This $s^3$ component couples with the non-zero components of the field strength tensor $F^3_{xy}$ and $F^3_{yx}$, creating a non-zero Lorenz force (Eq.~\eqref{Lorefor}), leading to the deviation of the components of the resulting beam towards its center. 

\textit{Analytical solution.}
The system of equations~\eqref{spidyn}-\eqref{spadyn} is nonlinear, but it can be solved analytically using Jacobi elliptic functions $\sn, \cn, \dn$. First of all, the spatial evolution $x(t),y(t)$ is fully controlled by the other degrees of freedom, but does not influence them, being absent from the equations \eqref{spidyn},\eqref{Lorefor}. We can therefore focus on the spin precession and the wave vector evolution first. We take $\alpha=\eta=\hbar=m=1$ for the analytical considerations in this section. 
The flat IFC passes through $q_x=1,q_y=0$ in this case.
We consider a particular component of the beam, which has an initial wave vector with $q_x(0)=1$ and an arbitrary $q_y(0)$. Along IFC, $q_x-1\sim -q_y^4$ (it is the absence of the quadratic term which makes the IFC "flat"), which justifies the consideration of a constant $q_x(0)=1$ in analytics, while the exact IFC contour is used in numerics in the last part of the discussion.

The equation for $s^3$ spin component is equivalent to the one for the nonlinear Duffing oscillator.  Its solution is 
\begin{equation}
    s^3(t)=-\frac{q_y(0)}{|q_y(0)|}\cn\left(\frac{t-T/4}{\sqrt{1-q_y^2(0)}},1-q_y^2(0)\right)
\end{equation}
This solution is shown in Fig.~\ref{fig2}(a) for two opposite initial wave vectors $q_y(0)$. The period is $T=4\sqrt{1-q_y^2(0)} K(1-q_y^2(0))$ ($K$ is the complete elliptic integral of the first kind). For small $q_y$, this period diverges logarithmically: $T\sim 4 \log(4/q_y)$.

Inserting this solution into~\eqref{Lorefor} (which becomes similar to Hamilton's equation for a harmonic oscillator with a time-dependent "spring constant") allows finding the explicit solutions for $q_x(t)$ and $q_y(t)$:
\begin{eqnarray}
    q_x(t)&=&q_x(0)\cos\theta(t)+q_y(0)\sin\theta(t)\\
    q_y(t)&=&-q_x(0)\sin\theta(t)+q_y(0)\cos\theta(t)\label{sicos}
\end{eqnarray}
where
\begin{widetext}
\begin{equation}
    \theta(t)=\frac{2\sqrt{1-q_y^2(0)}\arccos\left[\dn\left(\frac{t-T/4}{\sqrt{1-q_y^2(0)}},1-q_y^2(0)\right)\right]\sn\left(\frac{t-T/4}{\sqrt{1-q_y^2(0)}},1-q_y^2(0)\right)}{\sqrt{1-\dn^2\left(\frac{t-T/4}{\sqrt{1-q_y^2(0)}},1-q_y^2(0)\right)}}\label{phase}
\end{equation}
\end{widetext}
Two solutions for $q_y(t)$ are shown in Fig.~\ref{fig2}(b) (red -- $q_y(0)>0$, blue -- $q_y(0)<0$).

An important feature is that the points described by \eqref{sicos} remain on a circle $q_x^2(t)+q_y^2(t)=q^2$, because of which the group velocity projection $dy/dt$ is directly proportional to $q_y$ (that we have just found): $dy/dt=q_y(\hbar/m-\alpha/(\hbar q))$ thanks to $q$ in the last term being constant. The position along $y$ can therefore be determined as
\begin{equation}
    y(t)=\left(\frac{\hbar}{m}-\frac{\alpha}{\hbar q}\right)\int_0^t q_y(t')\,dt'\label{yint}
\end{equation}
An example of the behavior of $y(t)$ is shown in Fig.~\ref{fig2}(c) again for two opposite initial $q_y(0)$ (red, blue). The oscillatory behaviour of two opposite $q_y$ components leads to compensation, and the  center of mass of the full wavepacket, symmetric with respect to $q_y=0$ and linearly polarized,  does not oscillate in this regime, as already shown experimentally in \cite{polimeno2021experimental}. Only the width of the beam is affected by the gauge field.

While this integral cannot be calculated analytically, it can nevertheless be approximated using the properties of $q_y(t)$. This function exhibits symmetric regions of rapid variation, which cancel out, and two flat regions around the extrema at $t=0$ and $t=T/2$ (see Fig.~\ref{fig2}(b)). The value at $t=0$ is always the same: $q_y(0)$, while the value at $t=T/2$ changes, depending on $q_y(0)$.

Two limiting cases are of interest. First, for $q_y(0)\to 0$, one can find the limit $q_y(T/2)\to -3q_y(0)$. This difference in the amplitude of the extrema is partially compensated by the difference in their duration (ratio $2/3$), and the limit of the beam displacement is simply $y(t)\to q_y(0)t(\hbar/m-\alpha/(\hbar q))$, meaning that the gauge field has no effect in this case, except the inversion of the sign of the displacement. This limit is of a little practical significance, because the convergence with $q_y(0)$ is logarithmically slow and also because the study of this regime  requires exponentially long times due to $T\sim -\log(q_y(0))$. 

\textit{Canalization by gauge field.}
As $q_y(0)$ is increased, the contribution of the region at $t=T/2$ is reduced.
This can be understood via the decomposition provided by~\eqref{sicos}-\eqref{phase}. The extrema of the group velocity $dy/dt$ are located at the moments defined by the equation
\begin{equation}
    \frac{\partial q_y}{\partial t}=\frac{\partial q_y}{\partial\theta}\frac{\partial \theta}{\partial t}=0
\end{equation}
Due to the periodicity of the elliptic functions determining the phase according to~\eqref{phase}, the extrema at $t=0$ and $t=T/2$ are determined by the second coefficient, $\partial \theta/\partial t=0$. The additional narrow extrema correspond to $\partial q_y/\partial \theta=0$ and can be found as $\theta=\arctan(-q_x(0)/q_y(0))$. If this value exceeds the maximal possible value $\theta(T/2)$, the two negative minima disappear and a change of behavior occurs.
This is the second interesting point, where the positive and negative contributions to the position shift in~\eqref{yint} cancel each other exactly on each period $T$. It occurs around $q_y(0)\approx 0.46$, and at this point the gauge field leads to a complete suppression of any lateral drift. 
Between $q_y(0)=0$ and $q_y(0)\approx 0.46$, we observe therefore that the more the initial wave vector deviates from the main propagation direction, the more the lateral deviation is suppressed by the gauge field. This is illustrated in Fig.~\ref{fig2}(d) showing the final coordinate $y_f$ for a sufficiently long time $t=60 T_{max}$ (in order to have the oscillations of $y$ smaller than the average drift even for smallest wave vectors with the longest oscillation period) as a function of the initial wave vector $q_y(0)$ with the gauge field (magenta line) and without it (orange line). We also show the result for circular IFCs (no canalization -- black line) for comparison. The exact compensation point (where $y_f=0$) is visible as well. Beyond this point, the final deviation increases with $q_y(0)$.

In the absence of the gauge field, $y_f$ scales as $q_y^3$, in agreement with the series expansion of the group velocity~\eqref{vgy}, because the IFC is flat and not circular, so that $q\neq const$, giving $q_y/q-q_y\sim q_y^3$. The corresponding growth is much slower than for circular IFCs which have linear scaling $y_f\sim q_y$. In the presence of the gauge field one has a similar scaling $q^3$, but with a $q$-dependent coefficient (see Fig.~S1~\cite{suppl}).

\begin{figure}
    \centering
    \includegraphics[width=1.0\linewidth]{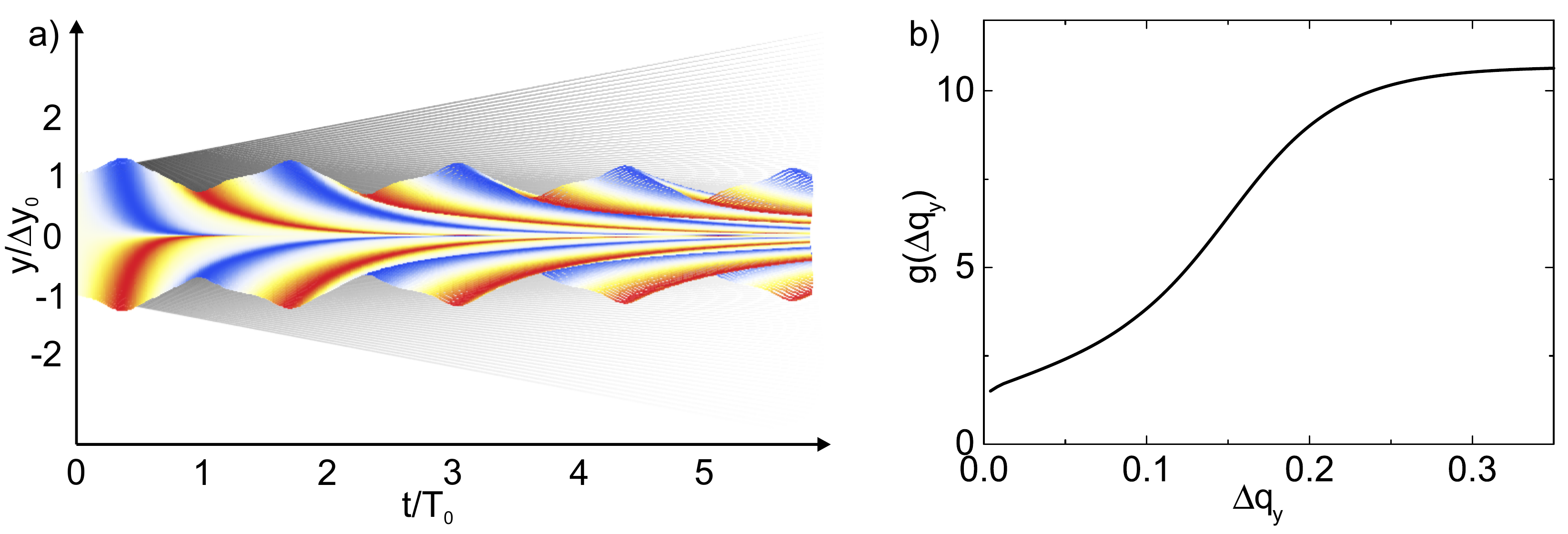}
    \caption{The effect of the gauge field. a) Transverse coordinate $y$ as a function of time $t$ for different starting points without gauge field (gray lines) and with gauge field (color -- $s^3$). b) The canalization effect of the gauge field $g$ for different initial beam size $\Delta q_y$.}
    \label{fig3}
\end{figure}

\textit{Finite size beam.}
In reality, a finite-size beam contains a lot of different wave vectors. We performed numerical simulations based on Eq.~\eqref{spidyn}-\eqref{spadyn} comparing two cases: with and without the gauge field. There is always a certain broadening due to the imperfect flatness of the IFC ($q_y^4$ term). The results are shown in Fig.~\ref{fig3}(a), where different trajectories for different initial conditions (with different positions $y(0)$ and wave vectors $q_y(0)$) are plotted together. In the case of a non-zero gauge field, the color corresponds to the value of the $s^3$ projection at each point, directly comparable with Fig.~\ref{fig1}(c). In the case of zero gauge field, the trajectories are plotted in gray (higher line density, equivalent to higher probability density, corresponds to darker gray). The difference between the spreads of the two beams is clearly visible: in the gauge field case, there is almost no visible spread, the beam is strongly canalized despite the non-completely flat IFC.

This effect of canalization by the gauge field can be evaluated quantitatively.
The final spread after a time $t=60T_{max}$ of a Gaussian beam depends on its initial 
width $\Delta q_y$:
\begin{equation}
    \Delta y=\frac{1}{\sqrt{2\pi\Delta q_y^2}}\int y_f^2(q_y(0))\exp\left(-\frac{q_y(0)^2}{2\Delta q_y^2}\right)\,dq_y(0)
\end{equation}
We define the extra canalization provided by the gauge field $g(\Delta q_y)$  as
the ratio between $\Delta y$ computed without the gauge field and $\Delta y$ computed with the gauge field. It is plotted in Fig.~\ref{fig3}(b). The value of $g(\Delta q_y)$ increases from $3.85$ for $\Delta q_y=0.1$ to $10.5$ for $\Delta q_y=0.3$, after which a saturation of the extra canalization is observed.

We therefore conclude that the effect of the gauge field allows gaining a factor 10 on the canalization provided by quasi-flat IFCs, when these IFCs are generated by tilted Dirac points associated with the polarization pseudospin. Our work shows that the optical spin Hall effect associated with the polarization degree of freedom and the spin-orbit coupling that can be described as a gauge field can play a very significant role in beam propagation beyond the predictions based solely on the group velocity at the IFCs.

\begin{acknowledgments}
This work was supported by the the ANR program "Investissements d'Avenir" through the IDEX-ISITE initiative 16-IDEX-0001 (CAP 20-25), the ANR projects MoirePlusPlus (ANR-23-CE09-0033) and HAWQ (ANR-25-CE47-7323)
\end{acknowledgments}

\bibliography{biblio}

\renewcommand{\thefigure}{S\arabic{figure}}
\setcounter{figure}{0}

\begin{figure}
    \centering
    \includegraphics[width=0.9\linewidth]{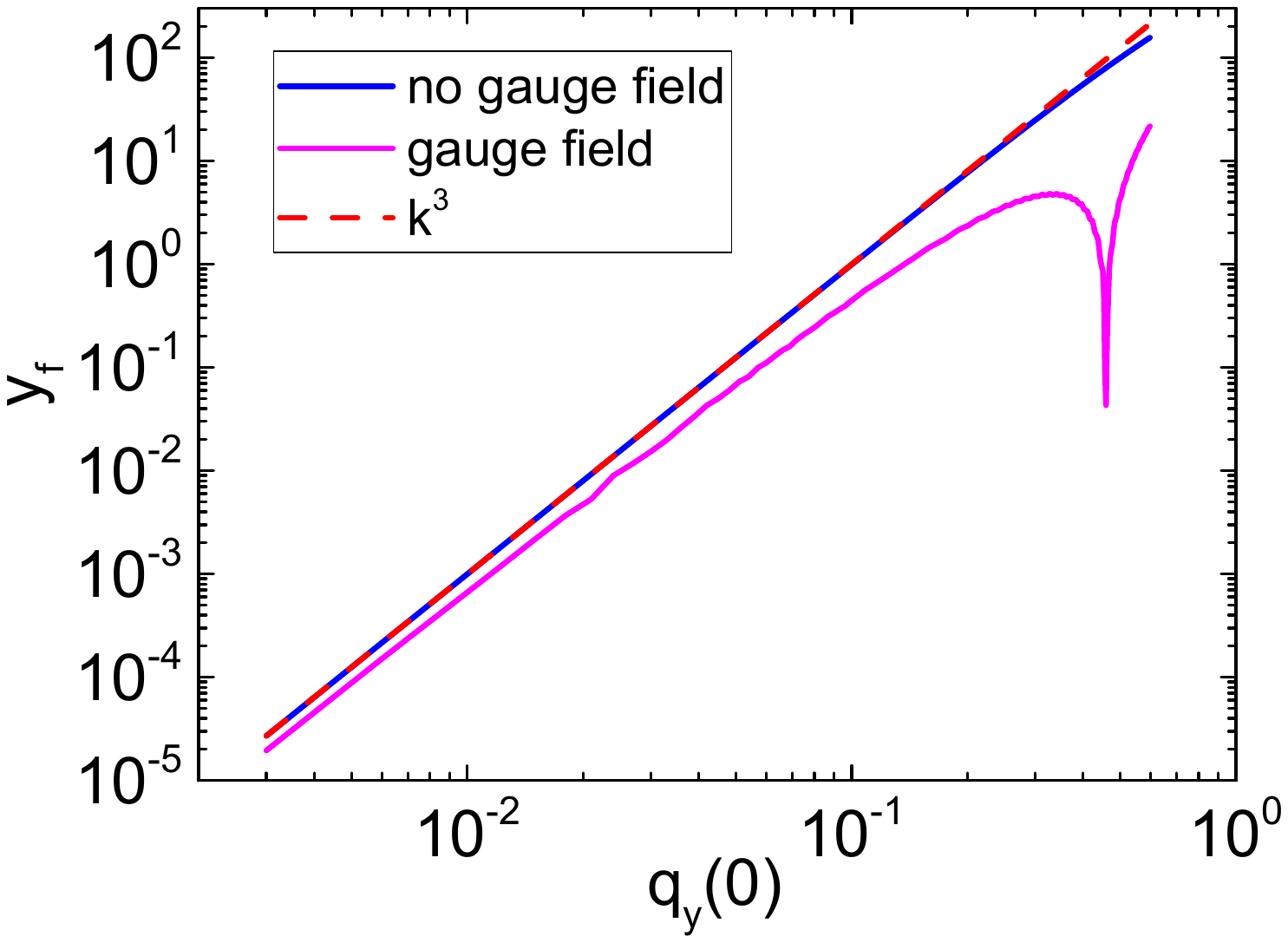}
    \caption{Log scale version of Fig. 2(d) from the main text, showing the final coordinate $y_f$ as a function of initial transverse wave vector $q_y(0)$.}
    \label{fig:my_label}
\end{figure}

\end{document}